\documentclass[12pt]{iopart}

\usepackage{wrapfig}
\usepackage{chemformula}
\usepackage{soul} 
\usepackage{braket}
\usepackage{amsmath}
\usepackage{amssymb}
\usepackage{bbold}
\usepackage{appendix}
\usepackage[utf8]{inputenc}
\usepackage[upgreek]{txgreeks}

\usepackage[sorting=none]{biblatex}
\usepackage[utf8]{inputenc}

\begin{document}

\title{Many-particle covalency, ionicity, and atomicity revisited for a few simple example molecules}

\author{Maciej Hendzel, Maciej Fidrysiak, and J\'{o}zef Spa\l{}ek}

\address{Institute of Theoretical Physics, Jagiellonian University, \L{}ojasiewicza 11, PL-30-348 Krak\'{o}w, Poland}
\ead{jozef.spalek@uj.edu.pl}
\vspace{10pt}
\begin{indented}
\item \today
\end{indented}

\begin{abstract}
We analyze two-particle binding factors of \ch{H2}, \ch{LiH}, and \ch{HeH+} molecules/ions with the help 
of our original exact diagonalization \emph{ab initio} (EDABI) approach. The 
interelectronic correlations are taken into account rigorously within the second quantization scheme for 
restricted basis of renormalized single-particle wave functions, i.e., with their size readjusted in the 
correlated state. This allows us to determine the many-particle covalency and ionicity factors in a 
natural and intuitive manner in terms of the microscopic single-particle and interaction 
parameters, also determined within our method. We discuss the limitations of those basic 
characteristics and introduce the concept of atomicity, corresponding to the Mott and Hubbard
criterion concerning localization threshold in many-particle systems. This addition introduces an atomic ingredient into the electron states and thus removes a spurious behavior of covalency with 
the increasing interatomic distance, as well as provides a more complete physical interpretation of bonding. 
\end{abstract}

\section{Introduction}

Determination of the microscopic nature of chemical bonding has been regarded as a problem of fundamental significance since the advent of quantum 
chemistry and solid state physics \cite{Piela2013, pauling1960nature, seitz1940modern}. The qualitative classification of the valence-electrons
state character as \emph{covalent}, \emph{ionic} or \emph{atomic} helps to rationalize their overall features and select a detailed approach to analyze their detailed electronic properties. In this respect, the role of interactions and associated with them interelectronic correlations is crucial in discussing the evolution of bonding from either atomic or ionic character to predominantly covalent or band states of valence electrons. The many-electron approaches, such as \textbf{C}onfiguration \textbf{I}nteraction (CI) \cite{Szabo1987}
and others \cite{Gimarc1979, Cooper2002}, are particularly well suited for this task.

In this work we follow a different route and employ \textbf{E}xact \textbf{D}iagonalization \emph{\textbf{A}b \textbf{I}ntio} (EDABI) method, combining the second-quantization formulation of quantum many-particle Hamiltonian with a concomitant readjustment of the single-particle wave functions in the correlated state of the system. This allows us to reinterpret some of the chemical bonding characteristics using concepts originating from condensed-matter physics, such as Mott-Hubbard localization. EDABI has been formulated in our group some time ago \cite{JS2000, Rycerz2001, JPCM2007} and analyzed extensively in the context of correlated states in small clusters and one-dimensional solid-state systems. Apart from providing rigorous description of selected properties, EDABI has supplied us with the evolution from the atomic- to a coherent-metallic state with decreasing interatomic distance. Also, modeling the metallization of molecular hydrogen solid has revealed a series of discontinuous first-order Mott-type transitions as a function of applied pressure \cite{BIBORSKI2015, Kadzielawa2017, Biborski2018}. The explicit question we would like to address here is to what extent the concepts essential extended to lattice quantum systems, such a \emph{Mottness} \cite{PRLSpal, Philips}, may also be qualitatively applicable to finite molecular systems. Answering this question forced us to reanalyze the meaning of the two-particle covalency and related to it ionicity factors by starting from an analytic form of many-particle wave function. We suggest that such analysis may be useful in practical treatment of bonding, here carried out in two-atom-molecule situation, to make the discussion analytic and thus provide a degree of clarity. 

The structure of the paper is as follows. In Sec.~\ref{sec:edabi_and_bonding} we summarize briefly the EDABI approach. In Sec.~\ref{sec:covalency_ionicity_atomicity} we reanalyze the bonding in \ch{H2}, \ch{HeH+}, and \ch{LiH} systems. We also discuss there validity of the concept of atomicity -- \emph{Mottness}, with the help of which we single out the \emph{resonant covalency} and atomicity factors. This discussion offers a resolution of the longstanding paradox of the increasing covalency with the increasing interatomic distance. Finally, in Sec. ~\ref{sec:overall_properties} and ~\ref{sec:outlook} we overview our approach.  Formal details and tabulated values of the calculated microscopic parameters as a function of interatomic distance are provided in Appendices~A-B.

\section{EDABI method and many-particle bonding}
\label{sec:edabi_and_bonding}

The EDABI method has been proposed by us and formulated in detail earlier \cite{JS2000, JPCM2007}. Below, we provide a brief summary of its main features, as this should by helpful in grasping the essence of our approach which will be needed in a subsequent interpretation of the results regarding many-particle covalency and ionicity, as well as the concept introduced by us of atomicity.

The starting point is the Hamiltonian containing all pairwise interactions in the second-quantized form is

\begin{align}
     \mathcal{\hat{H}} = & \epsilon_a \sum_i \hat{n}_{i\sigma} 
    + {\sum_{ij\sigma}}' t_{ij}\,\hat{a}^{\dag}_{i\sigma} \,\hat{a}_{j\sigma} + U \sum_{i} \hat{n}_{i\uparrow}\,\hat{n}_{i\downarrow} + \frac{1}{2} {\sum_{ij}}' K_{ij}\hat{n}_{i\sigma}\,\hat{n}_{j\sigma'}
     - \nonumber \\ 
    & \frac{1}{2}{\sum_{ij}}' J^H_{ij}  \left(\hat{\textbf{S}}_i \cdot \hat{\textbf{S}}_j-\frac{1}{4}
    \hat{n}_i\hat{n}_j\right) + \frac{1}{2} {\sum_{ij}}' {J}'_{ij}
    (\hat{a}^{\dagger}_{i\uparrow}\hat{a}^{\dagger}_{i\downarrow}\hat{a}_{j\downarrow}\hat{a}_{j\uparrow} + \mathrm{H.c.}) + \nonumber \\
    & \frac{1}{2} {\sum_{ij}}' V_{ij} (\hat{n}_{i\sigma}+\hat{n}_{j\sigma})(\hat{a}^{\dagger}_{i\bar{\sigma}}\hat{a}_{j\bar{\sigma}}+ \mathrm{H.c.}) + \mathcal{H}_{\text{ion-ion}},
    \label{Hamiltonian_eq}
\end{align}

\noindent
where H.c. denotes the Hermitian conjugation, $\hat{a}_{i\sigma}$ ($\hat{a}^\dagger_{i\sigma}$) are fermionic annihilation (creation) operators for state $i$ and spin $\sigma$, $\hat{n}_{i\sigma} \equiv \hat{a}^\dagger_{i\sigma} \hat{a}_{i\sigma}$, and $\hat{n}_i \equiv \hat{n}_{i\uparrow} + \hat{n}_{i\downarrow} \equiv
\hat{n}_{i\sigma}+ \hat{n}_{i\bar{\sigma}}$. The spin operators are defined as $\hat{S}_i \equiv \frac{1}{2} \sum_{\alpha\beta} \hat{a}^\dagger_{i\alpha} \sigma_i^{\alpha\beta} \hat{a}_{i\beta}$ with $\sigma_i$ representing Pauli matrices. The Hamiltonian contains the atomic and hopping parts ($\propto \epsilon_a$ and $t_{ij}$, respectively), the so-called Hubbard term $\propto U$; representing the intra-atomic interaction between the particles on the same atomic site \emph{i} with opposite spins, the direct intersite Coulomb interaction $\propto K_{ij}$, Heisenberg exchange $\propto J^H_{ij}$, and the two-particle and the correlated hopping terms ($\propto J_{ij}^\prime$ and $V_{ij}$, respectively). The last term describes the ion-ion Coulomb interaction which is adopted here in its classical form. 

We now proceed to definition of two-particle bonding in general situation and within the second-quantization representation. The $N$-particle state, $\ket{\Psi_N}$, may be expressed in terms of the $N$-particle wave function $\Psi(\textbf{r}_1, {\ldots}, \textbf{r}_N)$ and the corresponding field operators $\hat{\Psi}(\textbf{r}_1), 
{\ldots}, \hat{\Psi}(\textbf{r}_N)$ as 

\begin{align}
    \ket{\Psi_N} = \frac{1}{\sqrt{N!}} \int d^3r_1{\ldots}d^3r_N \Psi_N(r_1,{\ldots},r_N)
    \hat{\Psi}_1^{\dagger}(r_1)\ldots\hat{\Psi}_N^{\dagger}(r_N)\ket{0},
\end{align}

\noindent
with $\ket{0}$ being the universal vacuum  state in the Fock space (for pedagogical exposition see, e.g., \cite{Robertson}). Here we employ a short-hand notation $r_i \equiv (\textbf{r}_i,\sigma_i$), where $\sigma_i = \pm 1$ is the spin quantum number. We can revert this relation to determine the wave function $\Psi_N(r_1,{\ldots},r_N)$, namely

\begin{align}
    \Psi_{\alpha}(r_1,{\ldots},r_N) = \frac{1}{\sqrt{N!}} \braket{0|\hat{\Psi}_1(r_1){\ldots}\hat{\Psi}_N(r_N)|\lambda_{\alpha}}, 
    \label{wave:function}
\end{align}

\noindent
where $\ket{\lambda_{\alpha}}$ is the eigenstate for which the 
wavefunction $\Psi_{\alpha}$ i explicitly determined. For spin-conserving interaction, $\alpha=(\sigma_1,\sigma_2,{\ldots},\sigma_N)$ is fixed $N$-spin-configuration. In effect, we determine $\ket{\lambda_{\alpha}}$ states around ground eigenstate 
$\ket{\lambda_{\alpha}} \equiv \ket{\lambda_{\mathrm{min}}}$. Hamiltonian \eqref{Hamiltonian_eq}
is used to obtain the eigenstates which for two-electron \ch{H2} system are discussed analytically in \ref{appendix:h2_molecule}.  

Since we focus explicitly on the two-site systems, the set of microscopic parameters ($\epsilon_a$, $t$, $U$, $K$, $J$, $J'$, and $V$) is defined through integrals of orthogonalized single particle basis functions, $\{w_i(\textbf{r})\}$, used next to define the field operators in turn needed to construct the Hamiltonian~\eqref{Hamiltonian_eq}. They are defined briefly first \cite{JS2000, JPCM2007}, whereas the values of the microscopic parameters are defined in Appendix ~B, starting from the nonorthogonal basis set of adjustable Slater functions. Namely, the orthogonalized atomic (Wannier) orbitals for \ch{H2} molecule the $1s$ are defined via Slater orbitals $\{\psi_i(\textbf{r})\}$ in the usual manner

\begin{align}
    w_{i\sigma}(\textbf{r}) = \beta[\psi_{i\sigma}(\textbf{r})-\gamma
    \psi_{j\sigma}(\textbf{r})], 
\end{align}

\noindent 
where $\sigma \pm 1$ is spin quantum number $i \neq i = 1,2$, and the coefficients $\beta$ and $\gamma$ take the form

\begin{align}
\begin{cases}
    \beta = \frac{1}{\sqrt{2}}
    \sqrt{
    \frac{1+\sqrt{1-S^2}}{1-S^2}
    }, \\
    \gamma = \frac{S}{1+\sqrt{1-S^2}},
  \end{cases}
  \label{eq:beta_and_gamma}
\end{align}

\noindent
\begin{figure}[t]
    \centering
    \includegraphics[width=0.6\textwidth]{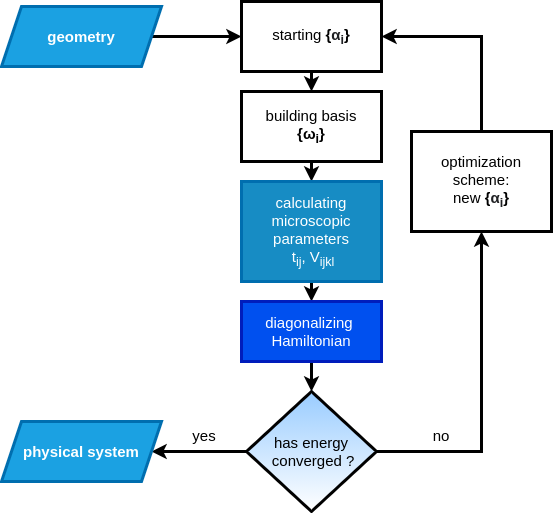}
    \caption{Flowchart of the EDABI method. The method is initialized by selection of a trial single-particle basis of wave functions, $\{w_i(\textbf{r})\}$, and subsequent diagonalization the resulting many-particle Hamiltonian. Optimization of the single-particle states leads to an explicit determination of the trial-wavefunction parameters, microscopic interaction and hopping parameters, ground-state energy, and explicit many-particle wavefunction, all in the correlated interacting state. For detailed discussion see main text.}
    \label{fig:flowchart}
  \end{figure}

\noindent
so that $\beta^2+\gamma^2 = 1$. The Slater orbitals 
$\psi_i(r) \equiv \sqrt{\frac{\alpha^3}{\pi}}\exp\left(-\alpha (\textbf{r}-\textbf{R}_i)\right)$, with $\alpha$ being the inverse size of the orbital, are to be readjusted during the ground-state-energy minimization in \emph{the correlated state}. Optimization over parameter $\alpha$ is motivated by the circumstance that the selected single-particle basis $\{\psi_i(\textbf{r})\}$ is never complete in the quantum mechanical sense and thus such a procedure allows for a better estimate of the ground-state energy. This allows for the orbital-size adjustment in the interacting environment of remaining particles. In brief, the standard procedure of determining the quantum-mechanical state of the system is inverted in the sense that we first diagonalize many-particle Hamiltonian for fixed values of the microscopic parameters and, subsequently, readjust the wave function (inverse size, $\alpha^{-1}$) in a recurrent fashion. The whole procedure is schematically illustrated by a flowchart composing Fig.~\ref{fig:flowchart}. 

The selected a single-particle basis for \ch{H2} is composed of four orthogonalized wave functions with indices $i= 1,2$ enumerating hydrogen atoms and $\sigma=\pm 1$ for each $i$. Thus, the truncated field operator takes the form

\begin{align}
    \hat{\Psi}(\textbf{r}) = \sum^2_{i=1} w_i(\textbf{r})\chi_{\sigma}(\textbf{r})\hat{a}_{i\sigma}.
\end{align}

\noindent
In that situation, the two-particle wave function is defined in accordance with Eq. \eqref{wave:function}, namely

\begin{align}
    \Psi_{\alpha}(\textbf{r}_1,\textbf{r}_2) = \frac{1}{\sqrt{2}} \braket{0|\hat{\Psi}_1(\textbf{r}_1)\hat{\Psi}_2(\textbf{r}_2)|\lambda_{\alpha}},
\end{align}

\noindent
where $\ket{\lambda_{\alpha}}$ is the eigenstate (expressed in second quantization representation). Note that this method of approach allows to determine both the ground-state and the lowest excited states for a single optimal value of $\alpha$. 

Note that here the subscripts 1 and 2 of $\hat{\Psi}(\textbf{r})$ contain both site and spin indices for brevity of notation. 
Parenthetically, one may generalize the above definition to the case of multiple ($n$) bonds (with $n \geq 1$) as

\begin{align}
    \Psi_a(\textbf{r}_1,{\ldots},\textbf{r}_{2n}) = \frac{1}{\sqrt{2n!}} \braket{0|\prod^n_{i=1}\hat{\Psi}_i(\textbf{r}_i)\prod^{2n}_{j=n+1}\hat{\Psi}_j(\textbf{r}_j)|\lambda^{(n)}_{\alpha}}.
\end{align}

\noindent
In this manner the double ($n=2$) and triple ($n=3$) bonds can be defined, albeit numerically only and in more complex situations, e.g., in the case of carbon-carbon bonds. This scenario is not addressed here. Instead, we focus on the covalency and ionicity, as well as introduce \emph{atomicity} + \emph{covalency} factor, all for selected two-electron systems. However, we discuss first the inherent paradox of the increasing covalency with the increasing interatomic distance.

\begin{figure}
    \centering
    \includegraphics[width=0.7\textwidth]{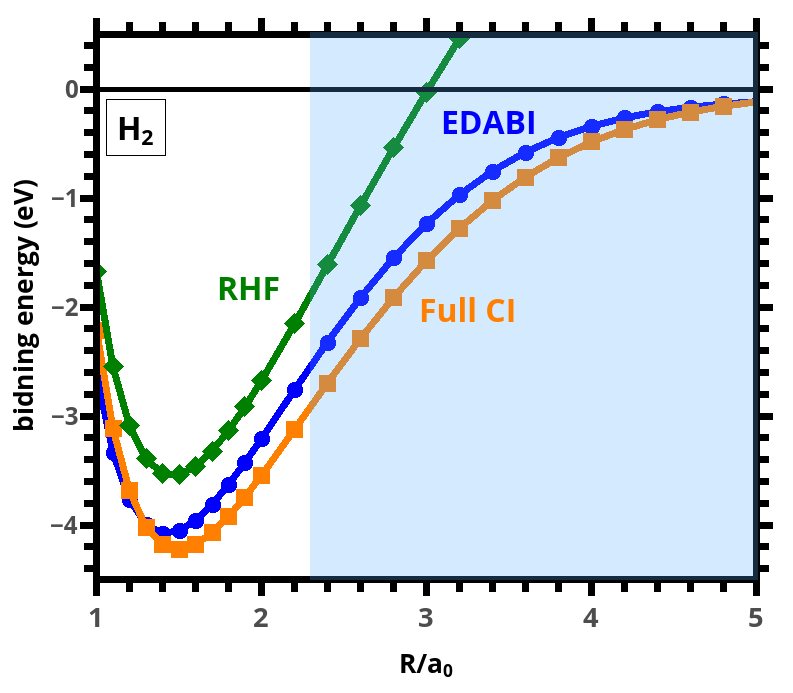}
    \caption{The \ch{H2} binding energy versus relative interatomic distance, calculated within EDABI and compared with restricted Hartree-Fock (RHF) and full configuration interaction (full CI) method. $a_0$ is the Bohr radius. EDABI yields lightly lower energy than full CI calculation at very small interatomic distance, $R$; this difference does not alter the main point of our qualitative discussion.}
    \label{fig:binding_h2}
\end{figure}

\section{Covalency, ionicity, and atomicity on examples}
\label{sec:covalency_ionicity_atomicity}

\subsection{Covalent bonding and ionicity in \ch{H2} case}

With the help of analysis presented in \ref{appendix:h2_molecule} for the \ch{H2} case, one can write down explicitly the two-electron wave function in the ground state for \ch{H2} molecule ($n=1$ case). The lowest-energy spin-singlet state is of the form

\begin{align}
     \Psi_0(\textbf{r}_1,\textbf{r}_2) = \frac{2(t+V)}{\sqrt{2D(D-U+K)}} \Psi_c(\textbf{r}_1,\textbf{r}_2)
     -\frac{1}{2}\sqrt{\frac{D-U+K}{2D}}\Psi_i(\textbf{r}_1,\textbf{r}_2)
     \label{coeff_wav}
\end{align}

\noindent
where the covalent ($\Psi_c$) and ionic ($\Psi_i$) components, read

\begin{align}
    \Psi_c(\textbf{r}_1,\textbf{r}_2) =& \left[ 
    w_1(\textbf{r}_1)w_2(\textbf{r}_2)+w_1(\textbf{r}_2)w_2(\textbf{r}_1)\right] \left[
    \chi_{\uparrow}(\textbf{r}_1)\chi_{\downarrow}(\textbf{r}_2)-\chi_{\downarrow}(\textbf{r}_1)\chi_{\uparrow}(\textbf{r}_2)\right], 
    \label{wav1} \\
    \Psi_i(\textbf{r}_1,\textbf{r}_2) =& \left[  
    w_1(\textbf{r}_1)w_1(\textbf{r}_2)+w_2(\textbf{r}_1)w_2(\textbf{r}_2)\right] \left[  
    \chi_{\uparrow}(\textbf{r}_1)\chi_{\downarrow}(\textbf{r}_2)-\chi_{\downarrow}(\textbf{r}_1)\chi_{\uparrow}(\textbf{r}_2)\right], 
    \label{wav2}
\end{align}

\noindent
with

\begin{align}
    D \equiv \sqrt{(U-K)^2+16(t+V)^2}.
    \label{D}
\end{align}

\noindent
The ratio of the coefficients in \eqref{coeff_wav} provides us with the relative ratio of covalency to ionicity in the ground-state
spin-singlet configuration. The spin-singlet part is the same in both Eq.~\eqref{wav1} and \eqref{wav2}. Explicitly, the covalency and ionicity coefficients (factors) are defined as

\begin{align}
    \gamma_c = \frac{16(t+V)^2}{16(t+V)^2+(D-U+K)^2}, 
    \label{eq:cov}
\end{align}

\noindent
and

\begin{align}
    \gamma_i = \frac{(D-U+K)^2}{16(t+V)^2+(D-U+K)^2},
\end{align}

\noindent
respectively, so that the condition $\gamma_i+\gamma_c = 1$ holds. The factor $\gamma_i$ asymptotically approaches zero in the limit of  large interatomic distance ($R\rightarrow \infty$), as expected. However, $\gamma_c \rightarrow 1$ for $R\rightarrow \infty$ which represents an \emph{unphysical} behavior \cite{An, Nat, Eu}. This last feature will be discussed in detail below.

\begin{figure}[htbp]
    \centering
    \includegraphics[width=0.7\textwidth]{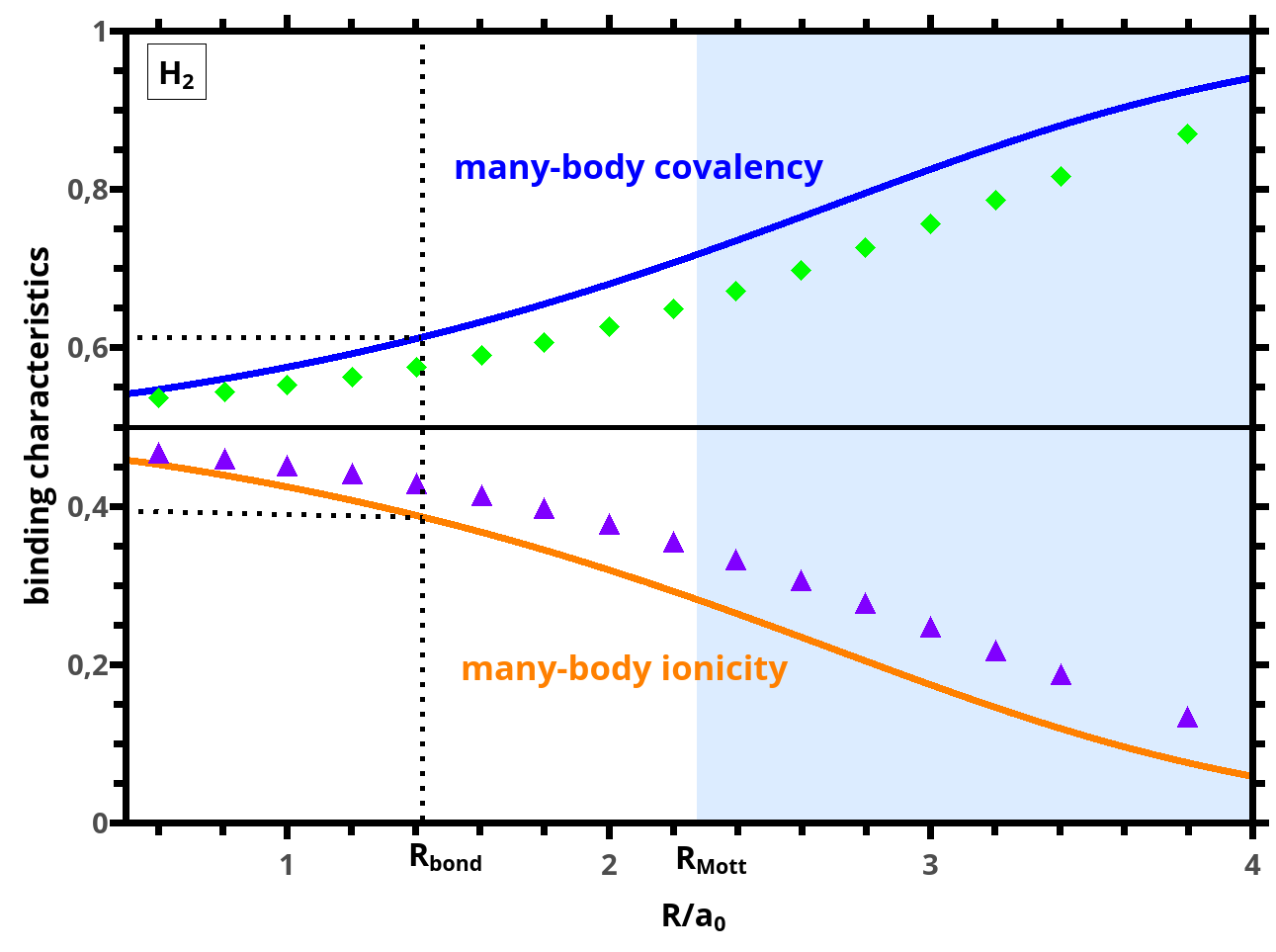}
    \caption{Two-particle covalency vs. corresponding ionicity for \ch{H2} molecule, calculated within EDABI method and compared with the results of Ref.~\cite{Pendas2018}. Shaded regime marks a gradual evolution towards \emph{atomicity}, as determined from the Mott-Hubbard criterion (see in the main text). Vertical dotted line marks equilibrium interatomic distance, whereas the horizontal dotted lines illustrate the dominant character of the covalency in that state (with the ratio $r = 1.43 \sim 2:1$)).}
    \label{fig:hcovion}
\end{figure}

The obtained formulas are interpreted as follows. First, the coefficients $\gamma_c$ and $\gamma_i$ in the 
wave function \eqref{coeff_wav} depend on all the interactions which are present in \eqref{Hamiltonian_eq}, 
i.e., they contain the effects electronic correlations. Second, the wave functions \eqref{wav1} and \eqref{wav2} take formally the 
Heitler-London form, but they are self-consistently optimized in the correlated state (their size $\alpha^{-1}$ is 
adjustable). Thus, the present formulation in its simplest from contains a semiquantitatively correct behavior in the 
large-$R$ limit, as is demonstrated explicitly in Fig.~\ref{fig:binding_h2}.  

\begin{figure}[t]
    \centering
    \includegraphics[width=0.7\textwidth]{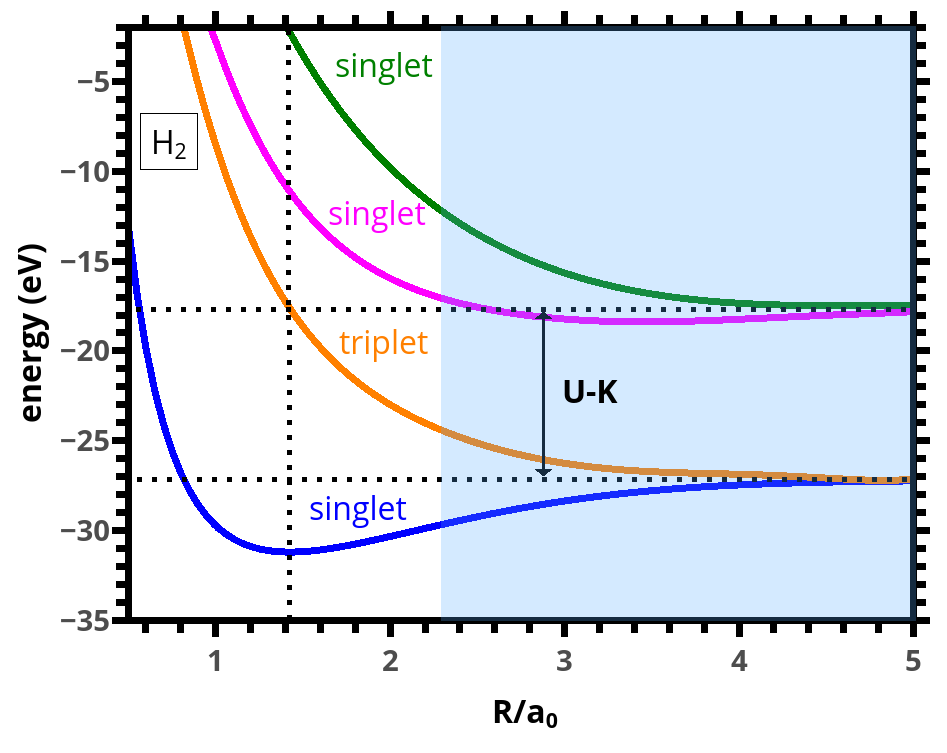}
    \caption{The lowest energy levels composed of three singlet and three triplet states, with the marked Mott regime and associated with it strong-correlation limit (shaded area). The scale $U-K$ represents the effective repulsive Coulomb interaction between electrons, i.e., the HOMO-LUMO splitting. The atomic character of the states increases with the increasing interatomic distance $R$.}
    \label{fig:singlet}
\end{figure}

In Fig.~\ref{fig:binding_h2}, we display the \ch{H2} binding energy and have compared our EDABI calculated value with the results of configuration interaction (CI) and restricted Hartree-Fock (RHF) analysis. Note difference between the results for small $R<R_{\mathrm{bond}}$ (at minimum), as EDABI method provides slightly lower energies compared to those of full CI. This behavior should not influence the subsequent discussion in the large-$R$ limit, which concerns us mainly here. Nonetheless, it is worth noting both CI and EDABI are variational approaches and perhaps our optimization of the wave-function size at small R is as important as the inclusion of the higher excited states. The binding energy is 
defined as $E_{\mathrm{bind}}=\lambda_5 - 2E_{H}$ ($\lambda_5$ is defined in Appendix ~A), where $E_H = - 1 \, \mathrm{Ry}$ is the energy of $1s$ state in atomic 
hydrogen.  Next, we define the bonding and ionicity as the corresponding ratios of coefficients in Eq.~\eqref{coeff_wav}, cf. Fig.~\ref{fig:hcovion}. We note that the covalency increases with the increasing interatomic distance at the expense of ionicity. However, this apparent inconsistency ignores the possibility of incipient atomicity of the \emph{Mott-Hubbard type}, i.e., the tendency towards localization of electrons on parent atoms with increasing R (called briefly \emph{the Mottness}). The Mott-type criterion for the localization of electron on \ch{H+} ion (i.e., formation of  renormalized atomic states) takes the form $2|t+V|/(U-K)=1$. This condition expresses the fact that the of bare kinetic  energy is then equal to the effective repulsive Coulomb interaction ($U-K$). In the strong correlation limit, the ratio
is below unity, meaning this repulsive interaction becomes predominant (note that in the strict atomic 
limit, $t+V \equiv 0$ whereas $U-K=1.25Ry$ then). The regime of strong-correlations (Mottness) is marked explicitly in Figs.~\ref{fig:hcovion} and \ref{fig:singlet}. It specifies a gradual evolution towards the atomic state.  Namely, the shaded area should be regarded as the regime with steadily increasing atomicity of the
electronic states with increasing $R$. Thus, the question of unphysically increased covalency for 
$R>R_{\mathrm{Mott}}$ is resolved in a natural manner as within the shaded area the covalency, $\gamma_c$, is composed of a sum of true (resonant) covalency $\bar{\gamma}_c \rightarrow 0$ and atomicity $\gamma_a \rightarrow 1$ as $R \rightarrow \infty$ (see the discussion below). 

\subsection{Correlation effects and incipient Mottness}

\begin{figure}[t]
    \centering
    \includegraphics[width=0.7\textwidth]{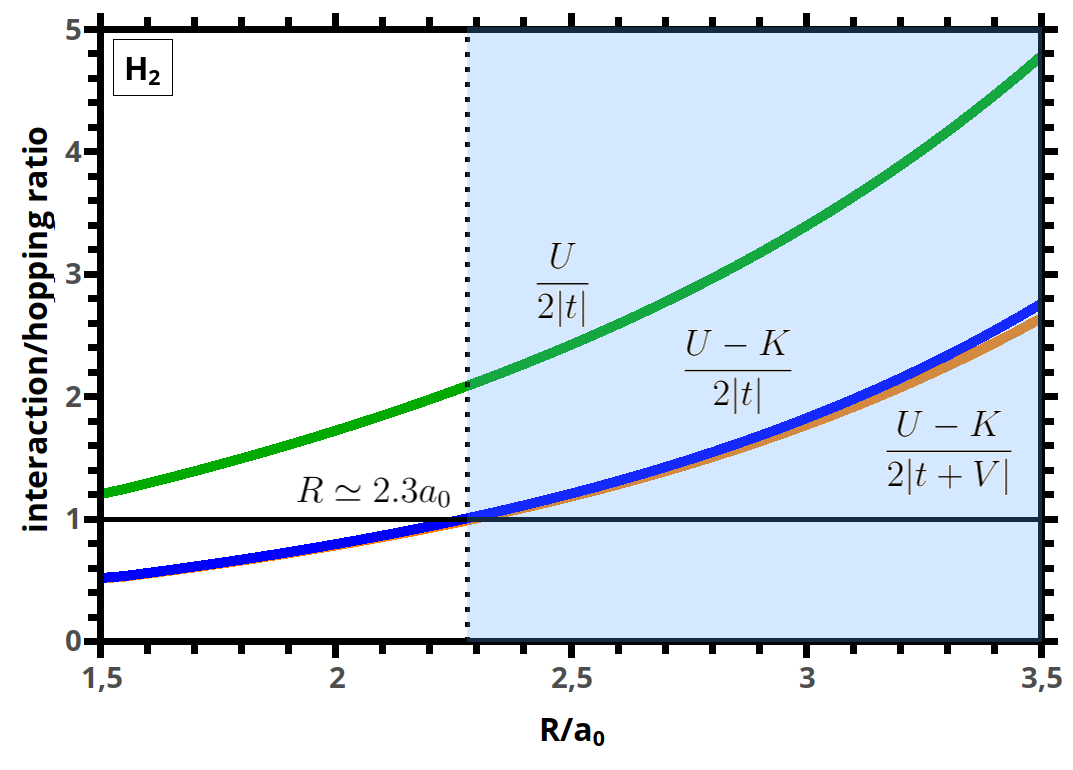}
    \caption{Characteristics of the \ch{H2} state. The condition $(U-K)/(2|t+V|)=1$ is the so-called Mott or Mott-Hubbard criterion for atomic localization which, in turn, determines the critical interatomic distance $R/a_0 \simeq 2.3$, representing the border of hatched area in Figs.~\ref{fig:hcovion} and \ref{fig:singlet}. The atomic character of electron wave function becomes gradually enhanced with increasing $R > R_{\mathrm{Mott}}$. The remaining curves have a supplementary character (see main text).}
    \label{fig:jvsj}
\end{figure}

The general meaning of the Mott (or Mott-Hubbard) effects is as follows. In condensed-matter physics the criterion takes the form  of $U \simeq W$ \cite{Mott}, where $W \equiv |\sum_{j(i)}t_{ij}|$ is the bare bandwidth. The transition takes the form of often discontinuous metal-insulator transition for odd integer number of relevant valence electrons per atom. In molecular system, such as \ch{H2}, the HOMO-LUMO splitting $\simeq U-K$ must overcome the effective interatomic hopping amplitude $W=2|t|$.  For Hamiltonian~\eqref{Hamiltonian_eq}, the Mott-Hubbard criterion takes then the form $r \equiv (U-K)/(2|t+V|) \simeq 1$ so that both the correlated hopping and intersite Coulomb interaction contribute, in addition to t and U. In the present situation, the criterion separates only qualitatively the regime of strong correlations ($r \geqslant 1$) from that with moderate to weak correlations ($r < 1$). Various versions of the criterion have been shown in Fig.~\ref{fig:jvsj}, depending on the theoretical model selected. Namely, the uppermost curve (in green) provides the criterion for the Hubbard model, which does not yield any Mottness point in the present situation. On the other hand, both the model with $V=0$ and the full model (represented by the starting Hamiltonian~\eqref{Hamiltonian_eq}), are almost identical and yield the critical interatomic distance for localization $R=R_{\mathrm{Mott}}\simeq 2.3 a_0$, i.e., well above the $R_{\mathrm{bond}} \simeq 1.43 a_0$. Note also that even for equilibrium distance $R=R_{\mathrm{bond}}$ the hopping/interaction ratio is about $\sim 0.5$, i.e., the electrons are moderately correlated. 

\begin{figure}[t]
    \centering
    \includegraphics[width=0.7\textwidth]{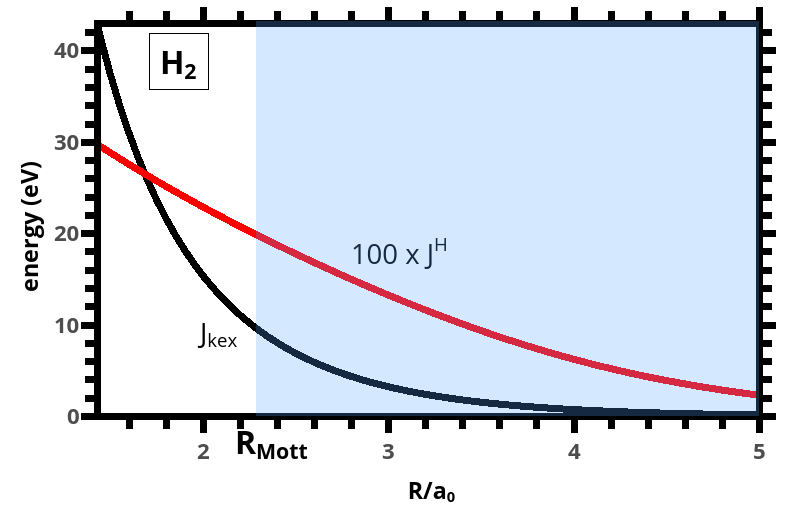}
    \caption{The antiferromagnetic kinetic exchange (superexchange) integral, $J_{\mathrm{kex}}$, calculated as a function of interatomic distance within EDABI approach. Superexchange dominates over its ferromagnetic Heisenberg correspondant, $J^H$, and provides the justification for the molecule spin-singlet configuration. The kinetic exchange originates from virtual resonant hopping of the electron between the atoms \cite{SpalS}.}
    \label{fig:jdoj}
\end{figure}

To complete the picture, we have also plotted in Fig.~\ref{fig:jdoj} the antiferromagnetic kinetic-exchange integral $J_{\mathrm{kex}} = 4(t+V)^2/(U-K)$ versus the direct (Heisenberg) ferromagnetic value $J^H$, both as a function of relative distance $R/a_0$. The situation is that $J_{\mathrm{kex}}>J^H$ for any distance $R$ and this is the reason for the spin-singlet configuration of \ch{H2} in the ground state. In brief, electrons hoping ("resonating") between the sites, possible only in the total spin-singlet state $\ket{\lambda_5}$, contribute essentially to the bonding.

\begin{figure}[htbp]
    \centering
    \includegraphics[width=0.7\textwidth]{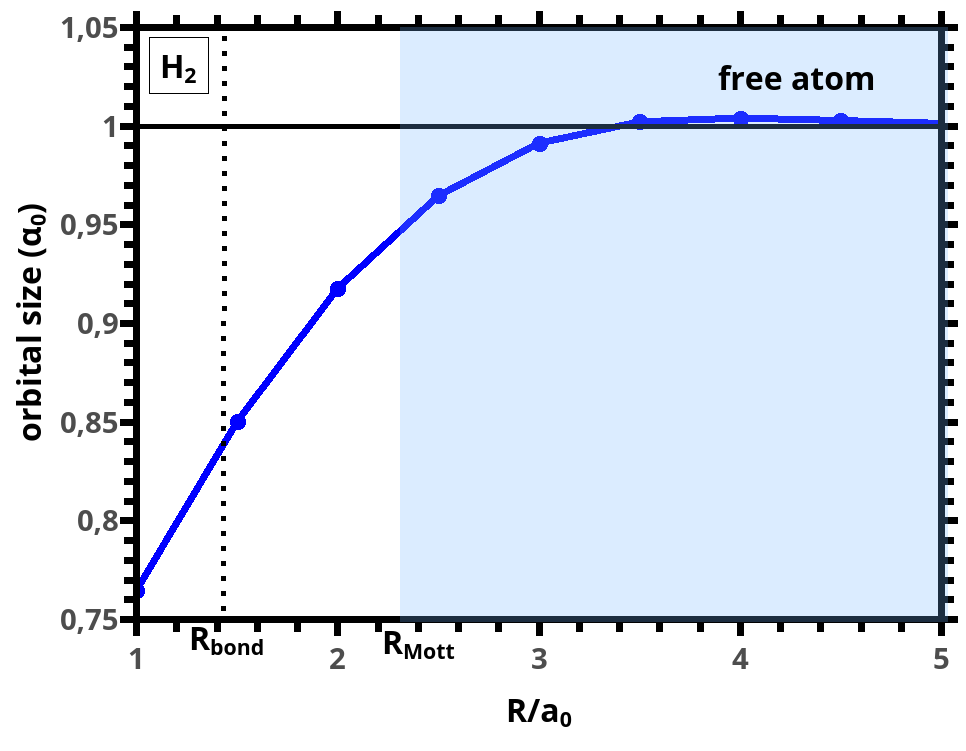}
    \caption{Renormalized $1s$ orbital size $\alpha^{-1}$ (in Bohr units $a_0$) vs. relative interatomic distance for the $\ch{H2}$ molecule. Note that after crossing the Mott-Hubbard point $R=R_{\mathrm{Mott}}$, $\alpha^{-1}$ approaches rapidly its atomic-limit value $\alpha^{-1}=a_0$. }
    \label{fig:alpha_h2}
\end{figure}

To verify the conceptual validity of the introduced Mott threshold 
for atomicity onset, we have plotted in Fig.~\ref{fig:alpha_h2} the Slater-orbital size $\alpha^{-1}$
as a function of $R$. Upon crossing the threshold $R_{\mathrm{Mott}}$, $\alpha^{-1}$ indeed approaches rapidly with the further increasing R the $1s$ atomic size value $a_0=0.53\,$\AA. Instead, the main physical process contributing to the bonding are the virtual process between the sites. In effect, the ionicity and covalency factors lose their principal meaning for $R \gg 2.3\,$\AA.

In conclusion, the dominant covalent character of \ch{H2} molecule has a well defined meaning for 
$R \simeq R_{\mathrm{bond}}$, as it is twice as large as the corresponding ionicity factor. 
However, this decomposition loses gradually its principal meaning as $R$ increases and crosses beyond $R_{\mathrm{Mott}} = 2.3 a_0$. The ground state energy evolves slowly, but steadily towards, the atomic-limit value. Note also that the Hartree-Fock analysis (cf. Fig.~\ref{fig:binding_h2}) provides 
unphysical results as this critical value of $R$ is crossed. This means that, in the regime of large interatomic 
distance, the role of correlation becomes essential. In effect, our analysis is applicable then and can be systematically extended numerically by, e.g., enriching the single-particle basis. It would be also of general interest to ask if those concepts could be tested quantitatively by putting \ch{H2} molecules on surfaces of other systems which would stretch the hydrogen-molecule size beyond the Mott-Hubbard threshold. Obviously, the analysis should then incorporate also the presence of the external surface potential of the substrate. However, this type of analysis goes beyond our goals here.

\subsection{Physical reinterpretation of atomicity, covalency, and ionicity: Resonant covalency}

\begin{figure}[htbp]
    \centering
    \includegraphics[width=0.8\textwidth]{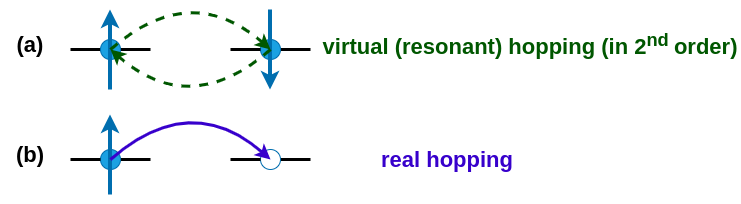}
    \caption{The virtual hopping processes that lead to the resonant covalency (a) and the real hopping, corresponding to the admixture of ionicity (b). For details see main text.}
    \label{fig:hopping}
\end{figure}

In order to provide a purely physical reinterpretation of covalency and ionicity we note that the form~\eqref{wave:function} of the covalent part contains sum of static products of the single-particle wave functions located on the sites 1 and 2 and their reverse; this is due to their indistinguishability in the quantum mechanical sense. On the contrary the coefficients $\gamma_c$ and $\gamma_i$ contain also virtual intersite processes depicted schematically in Fig.~\ref{fig:hopping}. In other words, the former factor contains a degree of atomicity in its static form, whereas the latter encompasses true dynamic virtual (hopping) processes of quantum-mechanical mixing. The question is how to separate those two factors into \emph{atomicity} and \emph{resonance covalency} parts in an analytic way. 

To answer this dilemma we propose its following resolution. The allowed local (site) states are $\ket{0,i}$, $\ket{\uparrow,i}$, $\ket{\downarrow,i}$, and $\ket{\downarrow,\uparrow,i}$, i.e., the empty, single occupied with spin $\sigma=\uparrow$ or $\downarrow$, or the double atomic occupancies. Therefore, using the following identities

    \begin{align}
        \ket{0,i}\bra{0,i} + \sum_{\sigma}\ket{\sigma,i}\bra{\sigma,i} + \ket{\uparrow, \downarrow, i}\bra{\uparrow, \downarrow, i} = \mathbb{I}, 
    \end{align}
    
    \noindent
    and its equivalent second-quantized from involving site occupancies

    \begin{align}
        \braket{(1-\hat{n}_{i\uparrow})(1-\hat{n}_{i\downarrow})} +
        \braket{\hat{n}_{i\uparrow}(1-\hat{n}_{i\downarrow})} + 
        \braket{\hat{n}_{i\downarrow}(1-\hat{n}_{i\uparrow})} + 
        \braket{\hat{n}_{i\uparrow}\hat{n}_{i\downarrow}} = 1.
    \end{align}
    
    \noindent
    Noting that the probability of empty atomic configuration is equal to that doubly occupied, i.e., physically corresponding to the electron-hole symmetry in condensed-matter systems, we obtain the formula for single-electron occupancy in the final form \cite{SpalHonig}
    
    \begin{align}
        \nu \equiv \sum_{\sigma} \braket{\hat{n}_{i\sigma}(1-\hat{n}_{i\bar{\sigma}})} = 1 - 2d^2.
    \end{align}
    
\noindent
Explicitly, we propose to decompose single-occupancy probability $\nu$ in the following manner

\begin{align}
    \nu \equiv a + c = 1 - 2d^2, 
\end{align}

\noindent
where $a$ is the atomicity, and $c$ is called the \textit{resonant (true) covalency}, and $d^2 \equiv 
\braket{\hat{n}_{i\uparrow}\hat{n}_{i\downarrow}}$ denotes atom double occupancy probability. Now, the resonant covalency describes the degree of mixing due to the virtual hopping admixture to the frozen (atomic) configuration (cf. Fig. \ref{fig:hopping}a). In the strong--correlation limit ($r>1$),  it can be defined as $c \equiv [|t-V|/(U-K)]^2$ and expresses the contribution of the processes (a) to the two-particle wave function in the second order \cite{HarrisLange, ChaoSpalek} as expressed by ratio of virtual (double hopping, forth and back) process to the Coulomb interaction change in the intermediate step. Therefore, the atomicity is evaluated as

\begin{align}
    a = \nu - c = 1 -2d^2 - c.
\end{align}

\noindent
In the equilibrium state of \ch{H2}, the resonant covalency reads  $c \simeq 0.8$, whereas atomicity $a \simeq 0.1$ is practically negligible. Conversely, with increasing $R$, $c$ decreases quite rapidly and approaches zero, whereas $a \rightarrow 1$, as anticipated.   

\begin{figure}
        \centering
        \includegraphics[width=0.7\textwidth]{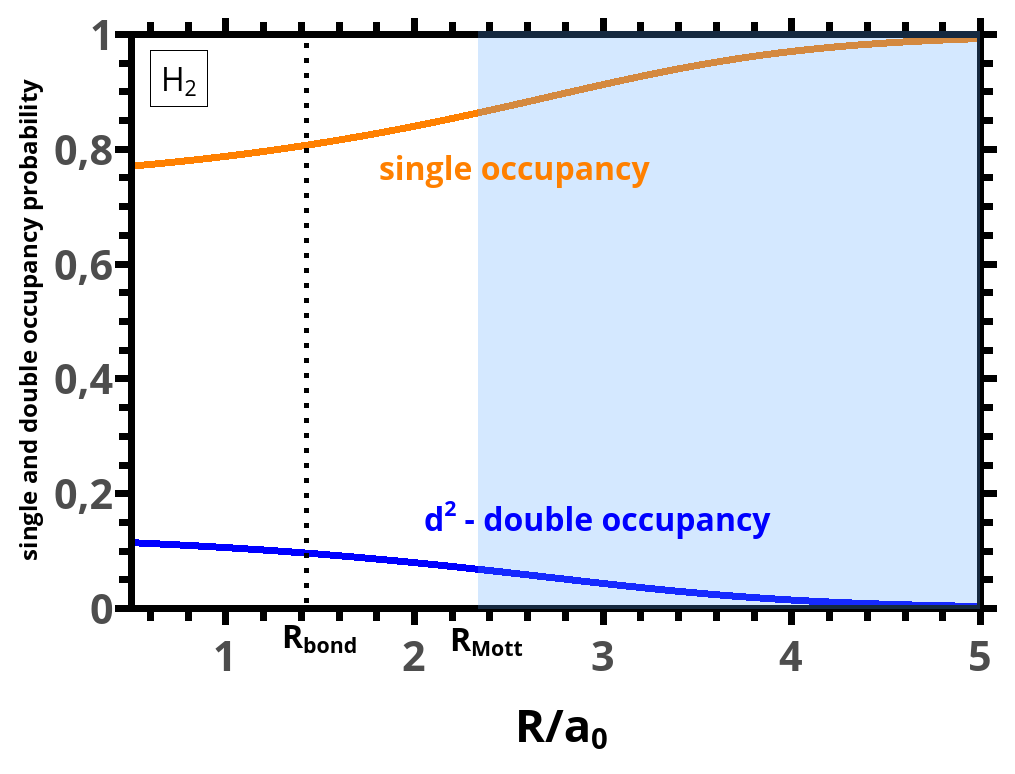}
        \caption{The atom double occupancy probability $d^2 \equiv \braket{\hat{n}_{i\uparrow}\hat{n}_{i\downarrow}}$ and simple occupancy $\nu$, both vs. $R/a_0$, calculated for \ch{H2} molecule using EDABI approach. Note the presence of inflexion point at $R=R_{\mathrm{Mott}}$, signaling the onset of gradually increasing single occupancy (the orange curve). The single occupancy $\nu$ contains both resonant covalency and atomicity, which cannot be separated from each other at this stage. For detailed discussion see main text.}
        \label{fig:occupancy}
    \end{figure}

Finally, in Fig.~\ref{fig:occupancy} we provide another characteristic containing atomicity, namely the $R$ 
dependence $d^2$. This double occupancy probability can also characterize the ionicity.
    The last formula shows that the atomicity is complete when $d^2 = 0$ and then $\nu = a = 1$ (i.e., for $R \gg R_{\mathrm{Mott}}$). In the other words, the customarily, defined by~\eqref{eq:cov} covalency, associated with the wave function \eqref{wave:function}, contains both atomicity and true covalency. For $R \gg R_{\mathrm{Mott}}$ it involves mainly atomicity with a small admixture of c and d$^2$. In this manner, the unphysical increase of $\gamma_c$ with increasing $R$ is resolved. In brief, fundamentally, we define the resonant (true) covalency \emph{c} as proportional to the inverse Mottness, i.e., 
    
    \begin{align}
        true \textrm{ } covalency = 1/Motness \quad \mathrm{or} \quad c \equiv 1/4r.
    \end{align}
    
    \noindent
    In conclusion, based on our analysis of \ch{H2} molecule we suggest that the covalency definition through the values of $\gamma_c$ is not conceptually precise, whereas the ionicity is properly accounted for either by $\gamma_i$ or $2d^2$. Additionally, in this way the \emph{redefined covalency} is complementary to the \emph{Mottness} and \emph{vice versa}.
    
\subsection{\ch{LiH} and \ch{HeH+} cases}

\begin{figure}[htbp]
    \centering
    \includegraphics[width=0.7\textwidth]{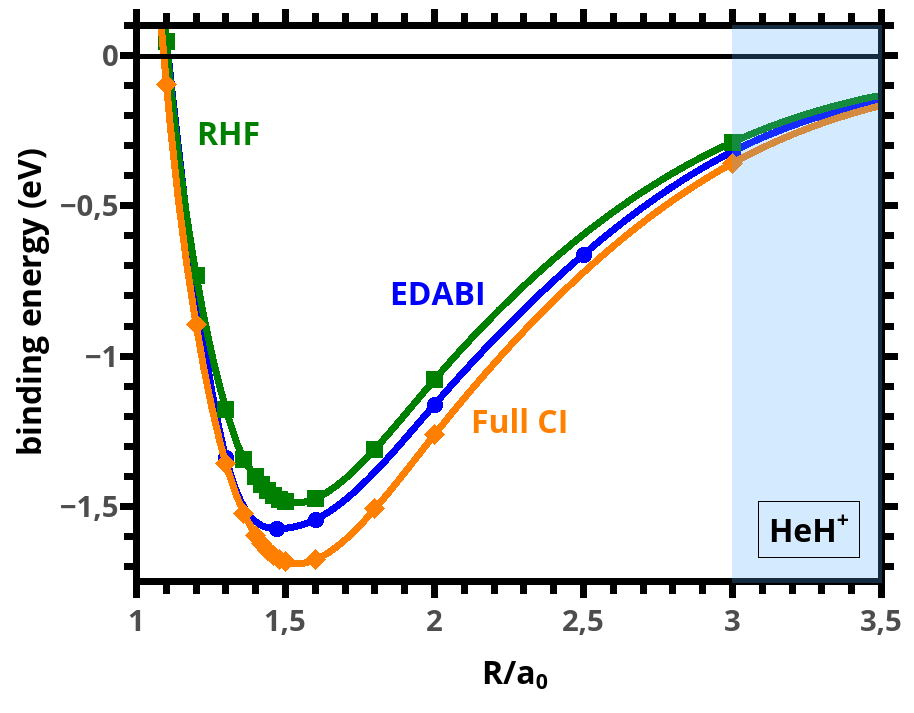}
    \caption{The \ch{HeH+} binding energy versus relative interatomic distance, obtained using EDABI method and compared with restricted Hartree-Fock (RHF) and full configuration interaction (full CI) approach. $a_0$ is the Bohr radius.}
    \label{fig:bind_heh}
\end{figure}

\begin{figure}[htbp]
    \centering
    \includegraphics[width=0.7\textwidth]{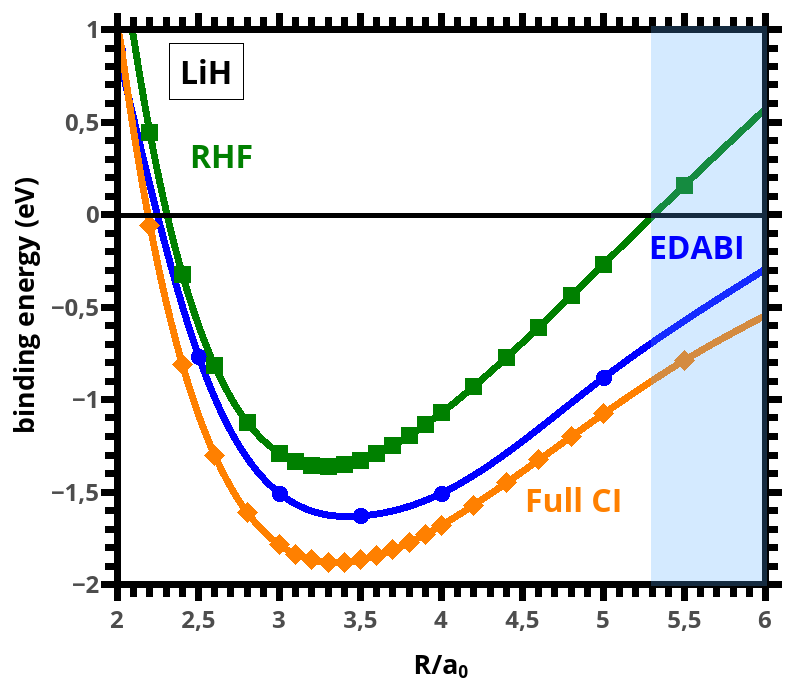}
    \caption{The \ch{LiH} binding energy versus relative interatomic distance, obtained using EDABI method and compared with restricted Hartree-Fock (RHF) and full configuration interaction (full CI) approach. $a_0$ is the Bohr radius.}
    \label{fig:bind_lih}
\end{figure}

We now apply the concepts introduced above for \ch{H2} molecule to the cases of \ch{LiH} and \ch{HeH+}. In Figs.~\ref{fig:bind_heh} and \ref{fig:bind_lih} we display the binding energies versus interatomic distance for \ch{HeH+} molecular ion and \ch{LiH} molecule, respectively. In the former case, the two $1s$ electrons are regarded as core electrons. Effectively, \ch{LiH} is regarded as a molecule composed of one $2s$ electron due to Li and $1s$ electron due to H, with their orbitals adjustable when the interactions are included. Qualitatively, the character of these curves is similar to those of \ch{H2}, depicted in Fig.~\ref{fig:binding_h2}. The quantitative factors are different though and, in particular, the bond length is slightly larger than that in \ch{H2} case.

\begin{figure}[htbp]
    \centering
   \includegraphics[width=0.7\textwidth]{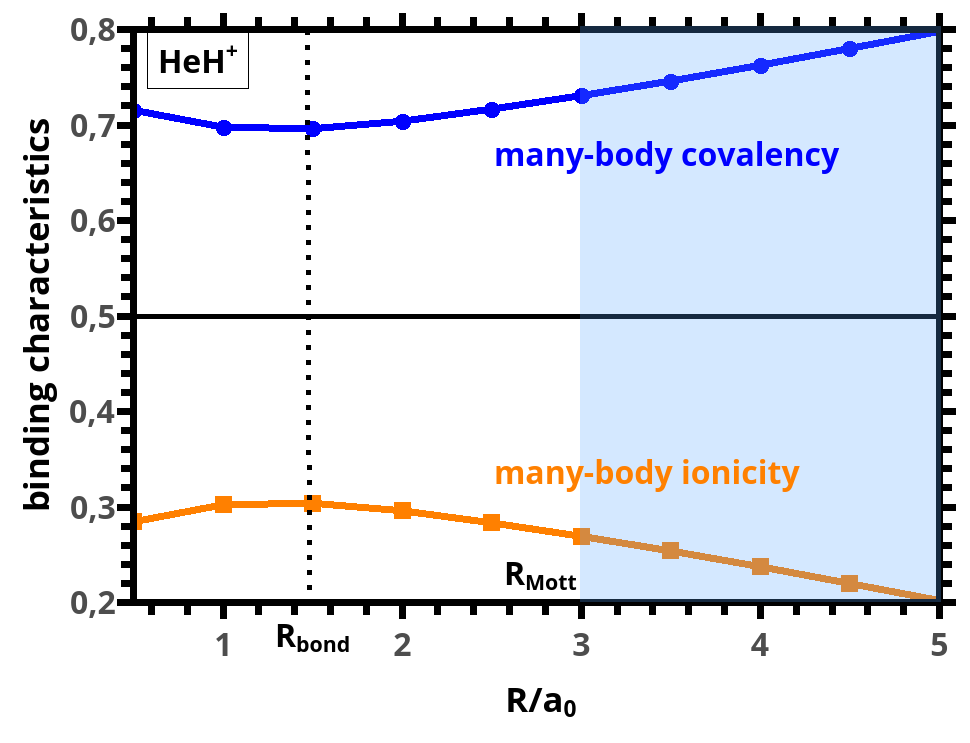}
    \caption{Many-body covalency vs. many-body ionicity for \ch{HeH+} molecule. The behavior is quite similar to that for \ch{H2} molecule (cf. Fig ~\ref{fig:hcovion}).}
    \label{fig:hehcovion}
\end{figure}

\begin{figure}[htbp]
    \centering
    \includegraphics[width=0.7\textwidth]{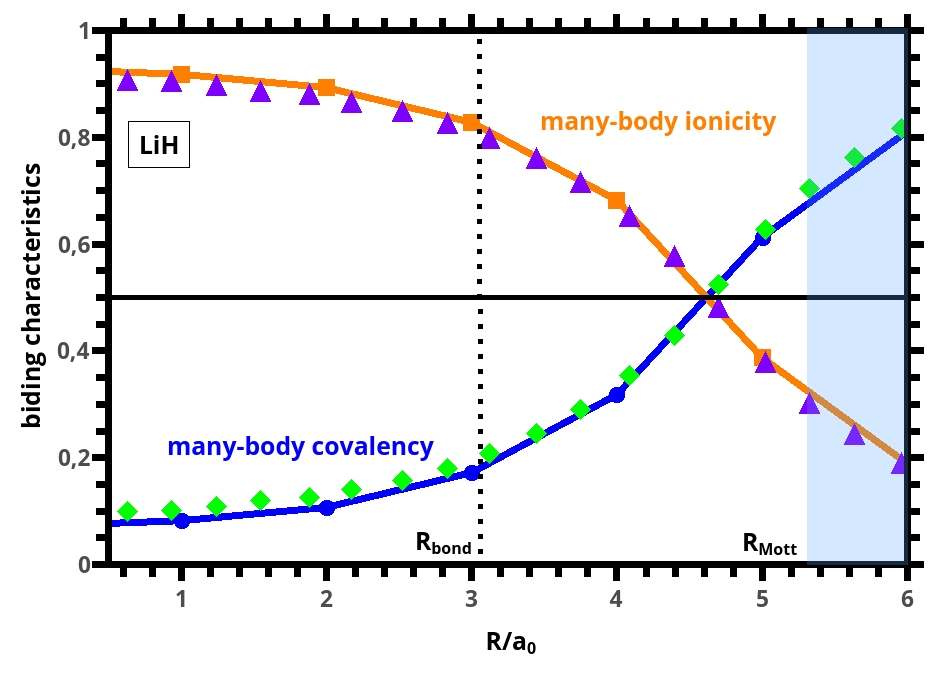}
    \caption{Many-body covalency vs. many-body ionicity for \ch{LiH} molecule. The 
    points are taken from \cite{Pendas2018} for comparison. The covalency shows the same type of the unphysical R-dependence as in the case of \ch{H2}.}
    \label{fig:lihcovion}
\end{figure}

To characterize further those two cases we have plotted in Figs.~\ref{fig:hehcovion} and \ref{fig:lihcovion} the covalency and ionicity factors for those two systems, respectively. Note that for \ch{LiH} the ionicity is predominant in a wide range of $R$, whereas the opposite is true for \ch{HeH+}. The difference arises from the circumstance that in \ch{LiH} case the orbital size of $2s$ electron is decisively larger and has a higher energy leading to predominantly ionic configuration $ \sim Li^{+0.9}H^{-0.9}$. In \ch{HeH+}, molecular ion the bonding is largely covalent due to the fact that both two $1s^2$ He electrons hop (mix) with the \ch{H+} state with no electrons in the corresponding $1s$ state.

\begin{figure}[htbp]
    \centering
    \includegraphics[width=0.7\textwidth]{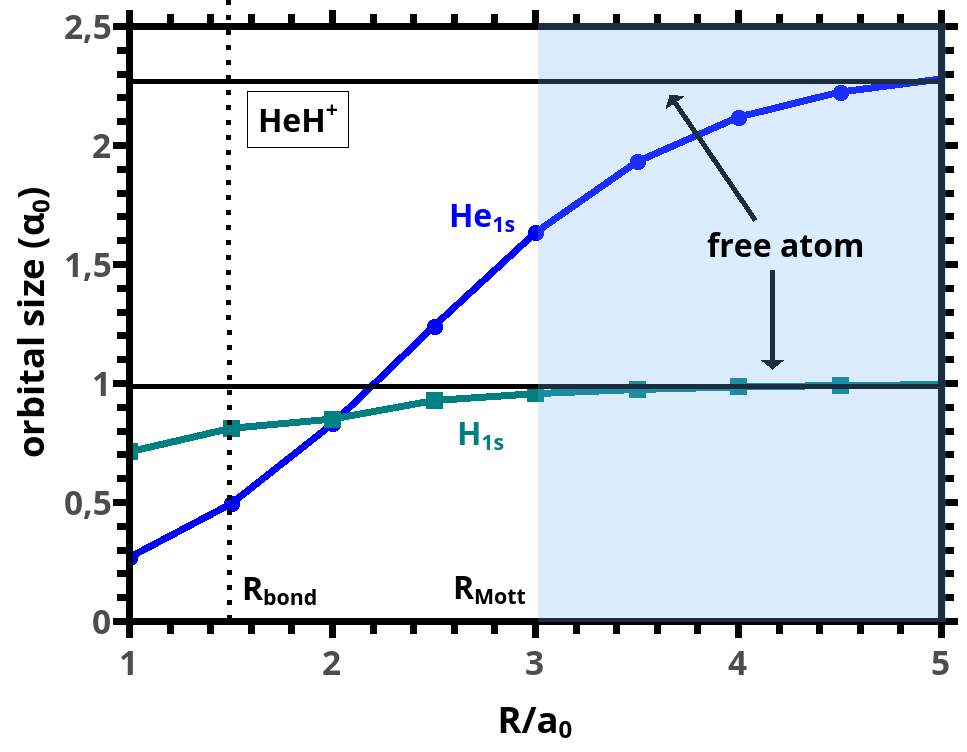}
    \caption{Atomic orbital size for He $1s$ and H $1s$ orbitals in \ch{HeH+} molecular ion. Note that the Mott-type boundary has been drawn for the $1s$ states of He as this reached first upon increasing R.}
    \label{fig:orb_size_heh}
\end{figure}

\begin{figure}[htbp]
    \centering
    \includegraphics[width=0.7\textwidth]{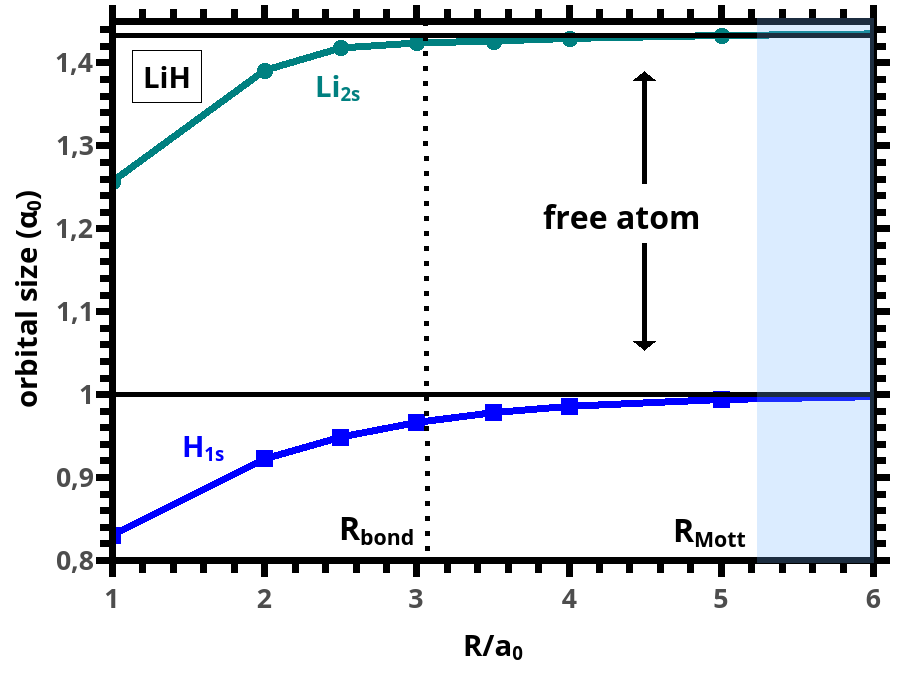}
    \caption{Atomic orbital size for H $1s$ and Li $2s$ orbitals in \ch{LiH} molecule.Note the strong renormalization of atomic-orbital sizes, as well as a rapid convergence to the atomic values above $R=R_{bond}$, the latter being for the ionic bonding. }
    \label{fig:orb_size_lih}
\end{figure}

One specific feature of those two systems should be noted, which is illustrated in Figs.~\ref{fig:orb_size_heh} and \ref{fig:orb_size_lih}, where the optimized sizes of the relevant orbitals has been shown. Namely, the size of $1s$ orbital of the He and $2s$ orbital of Li are strongly renormalized, the former largely expanded, whereas the latter contracted. The principal cause of this effect is the electronic correlation induced by the strong intraatomic (Hubbard) interaction $\sim U$. As this interaction in He is reduced by the flow of electron to the \ch{H+} site, it is not so in the case of Li, where presence of the hydrogen electron strongly enhances the role of the interaction. In spite of those differences, both systems exhibit similar span of covalency regime. On the contrary, the incipient Mottness appears for larger distance $R_{\mathrm{Mott}}$ and this is presumably due to a larger renormalized-orbital size for Li. As can be seen from literature \cite{HeHcovalency} and from our results here, \ch{HeH+} is largely covalent and the whole analysis of a and c factors can be repeated here without any qualitative difference. 

\clearpage
\newpage

\clearpage
\newpage

\clearpage
\newpage
\begin{table}[htbp]
\caption{Binding energy for \ch{H2}, \ch{HeH+}, and 
\ch{LiH} molecules (in eV).}
\centering
\label{tab::binding}
\begin{tabular}{|c|c|c|c|}
\hline
Method & \ch{H2} & \ch{HeH+} & \ch{LiH} \\ \hline
EDABI   & -4.0749 & -1.5803 & -1.6537 \\ \hline
Full CI & -4.3824 & -1.6849 & -1.8846 \\ \hline
RHF     & -3.5963 & -1.4839 & -1.3616 \\ \hline
Reference values   & -4.3821\cite{Kolos1964} & -2.0542\cite{Wolniewicz1965} &   -1.3606 \cite{KaroOlson} \\ \hline
\end{tabular}
\end{table}
In Table \ref{tab::binding} we display the binding energies of the molecules \ch{H2}, \ch{HeH+}, 
\ch{LiH}, regarded here as testing ground of our approach. For that reason we compare the obtained results 
with those deducted from other methods and with use of a richer single-particle basis. Even though our 
results are quantitatively not too accurate, they are obtained with simplest nontrivial basis, i.e., $1s$ states only for 
\ch{H2} and \ch{HeH+} cases, and with addition of $2s$ states on \ch{Li} the \ch{LiH} case. The 
EDABI results can be improved in a straightforward manner at the expense of computational resources. However in such a 
situation our following next discussion of bonding would be purely numerical. In other words, we accept the
lower accurateness of $E_G$ value within our method to allow for the analytic character of the subsequent 
discussion. One can find more accurate value of ground state energy for \ch{HeH+} \cite{PachuckiHeH}.

\section{Overall properties}
\label{sec:overall_properties}

\begin{wrapfigure}{r}{0.6\textwidth}
    \centering
    \includegraphics[width=0.6\textwidth]{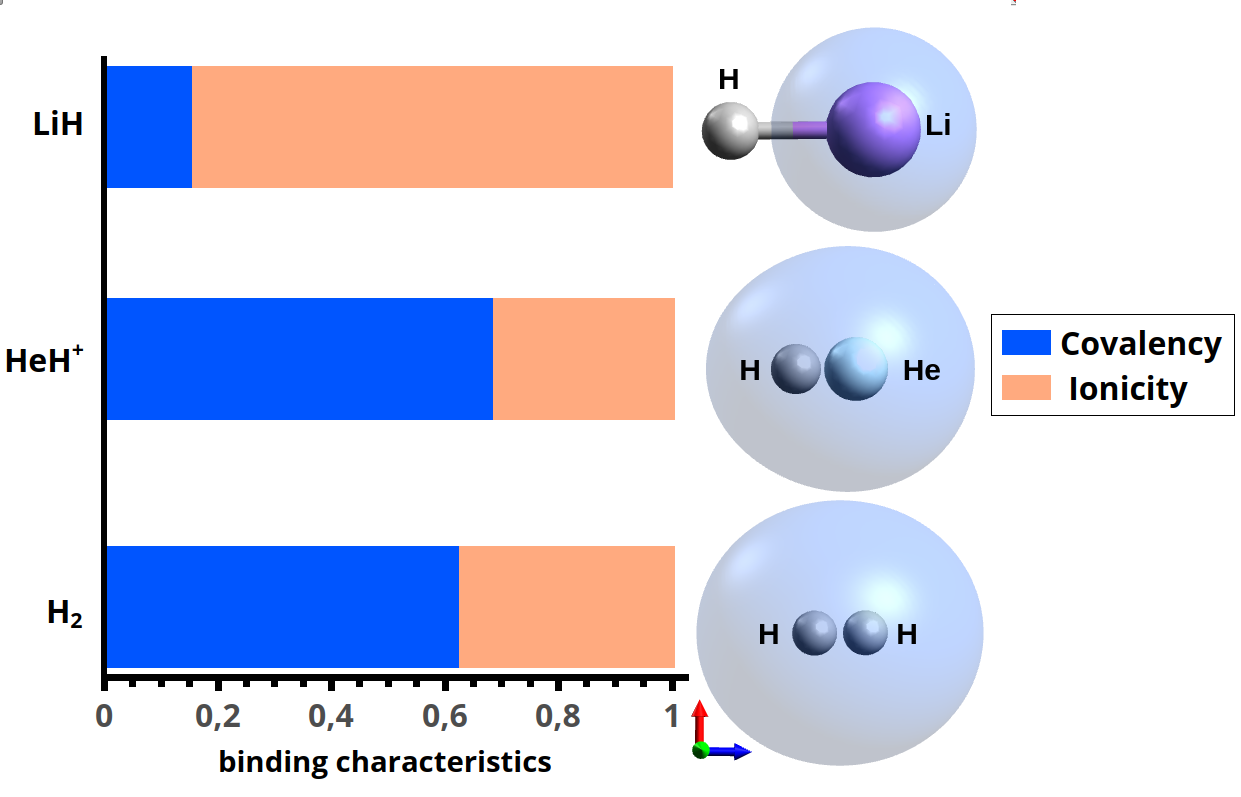}
    \caption{Schematic representation of the molecular orbitals for (isosurface probability density cut = 0.02) on the right, expressing relative covalency and ionicity contributions on the left.}
    \label{fig:covion}
\end{wrapfigure}
We now compare results for those three model systems qualitatively. First, in Fig.~\ref{fig:covion} we 
plot relative contributions of the covalency and ionicity factors for the two-particle ground state (left),
as well as a schematic size of the molecular orbitals relative to their original (atomic) size (right). The
atomicity factor is not quantified at this stage. In the first two of them, the dynamics is solely due to 
1s electrons, whereas in the \ch{LiH} case the $1s^2$ configuration of electrons is frozen on Li and the 
whole dynamics is due to $1s$-$2s$ H-Li mixing and the corresponding interactions. This is the reason why 
\ch{LiH} is largely ionic, whereas the remaining two are predominantly covalent, as illustrated in 
Fig.~\ref{fig:lihcovion}. One sees that the covalency in \ch{HeH+} is larger than that for \ch{H2} molecule, a rather unexpected intuitively result.

\clearpage
\newpage
\begin{table}[htbp]
\caption{The interatomic distance corresponding to the Mott boundary regime calculated for \ch{H2}, \ch{HeH+}, and \ch{LiH} molecules.}
\centering
\begin{tabular}{|c|c|}
\hline
  System                   &  Mott boundary ($a_0$)   \\
  \hline
\ch{H2}   & 2.3 \\ \hline
\ch{HeH+} & 3.0   \\ \hline
  \ch{LiH}  & 5.3 \\ \hline
\end{tabular}
\label{Mott-table}
\end{table}

A separate discussion should be concerned with other overall properties of the systems studied. In Table~\ref{Mott-table} the Mott (or Mott-Hubbard) critical distance $R_{\mathrm{Mott}}$ (in the units of $a_0$) is provided. This distance should be compared with the bond length $R_{\mathrm{bond}}$ calculated (cf. Table~\ref{Tab::bondlength}) according to three independent methods: EDABI, full configuration-interaction (CI), and restricted Hartree-Fock (RHF) methods, respectively. We see that in each case $R_{\mathrm{bond}}$ is decisively lower than $R_{\mathrm{Mott}}$. This means that the Mott-type boundary can be crossed only in the situation when the molecules are further apart, i.e., obtained artificially when, e.g., they are placed on surfaces with an external force elongating them. Particularly favorable situation occurs when molecules are placed in the environment with a large dielectric constant, as then interaction weakens and the bond length increases. Clearly, then the whole analysis must be revised and a realistic configuration with inclusion of appropriate external (surface) potential. We believe that the essential features of our analysis should survive when the molecule is placed in such environment, i.e., in a potential stretching equally both atoms.

\begin{table}[htbp]
\caption{Bond length for \ch{H2}, \ch{HeH+}, and 
\ch{LiH} molecules (in units of $a_0$).}
\centering
\label{tab::bond}
\begin{tabular}{|c|c|c|c|}
\hline
Method & \ch{H2} & \ch{HeH+} & \ch{LiH} \\ \hline
EDABI   & 1.430 & 1.469 & 3.382 \\ \hline
Full CI & 1.501 & 1.497 & 3.298 \\ \hline
RHF     & 1.450 & 1.493 & 3.208 \\ \hline
Reference values   & 1.398 \cite{Kolos1964}& 1.463 \cite{Wolniewicz1965} & 3.015 \cite{LiHAcc}  \\ \hline
\end{tabular}
\label{Tab::bondlength}
\end{table}

In Table~\ref{tab::binding} the binding energies are listed and compared with those from other methods. These numerical results present probably the weakest point of our EDABI method, since the corresponding values obtained are not very accurate. Nevertheless, we do not consider our method as a practical computing tool. Instead our main aim here was to extend, albeit at best in a semiquantitative manner, the basis for multi-electron covalency and ionicity, enriched by the concept of atomicity, all induced by the electronic correlations. Obviously, the approach can be extended in a straightforward manner by enlarging the single-particle basis and applied for the systems with larger atoms. Both of these factors have been considered by us before \cite{JS2000, JPCM2007} for model systems, with one limitation, that we have not analyzed there the bonding properties. This analysis should be explored further along the lines discussed here. 

\begin{table}[htbp]
\caption{Calculated EDABI values of the equilibrium parameters for \ch{LiH}, \ch{HeH+}, and \ch{H2} molecules.}
\centering
\begin{tabular}{|c|c|c|c|}
\hline
\textbf{Parameter} & \textbf{\ch{LiH}} & \textbf{\ch{H2}} & \textbf{\ch{HeH+}}   \\ \hline
$E_G$      (eV)        & -217.45      & -31.198     &      -79.675     \\ \hline
$U_{1s\mathrm{Li}}$   (eV)     & 49.062        & N/A           &          N/A     \\ \hline
$U_{1s\mathrm{H}}$   (eV)      & 18.559        & 22.490       &       19.592       \\ \hline
$U_{1s\mathrm{He}}$   (eV)      & N/A       & N/A      &       14.925        \\ \hline
$U_{2s\mathrm{Li}}$   (eV)     & 12.784        & N/A           &        N/A       \\ \hline
$t$  (eV)             & -21.150       & -9.9049      &      -15.674         \\ \hline
$K$  (eV)            & 14.368        & 13.007       &    11.374           \\ \hline
$\alpha_{1s\mathrm{H}}$   ($a_0^{-1}$)   & 1.035       & 1.194       &  1.240    \\ \hline
$\alpha_{1s\mathrm{He}}$   ($a_0^{-1}$)   & N/A       & N/A       &  2.095   \\ \hline
$\alpha_{2s\mathrm{Li}}$   ($a_0^{-1}$)   & 1.329        & N/A           &   N/A   \\ \hline
\end{tabular}
\label{Tab:eqstate}
\end{table}

Finally, in Table~\ref{Tab:eqstate} we list the most important microscopic parameters in the equilibrium state. A more detailed analysis of those is presented in Appendix ~B. The values of Coulomb-interaction parameters will be reduced by the dielectric constant factor if system under consideration is placed on surface of an insulating material. This should rescale all the parameter values accordingly. 

\section{Outlook}
\label{sec:outlook}

The reason for selecting the three systems analyzed here is caused by the circumstance that \ch{HeH+} is strongly covalent, \ch{LiH} strongly ionic, and \ch{H2} can be placed in between them. On example of the last of them our novel concept of atomicity and resonant covalency have been proposed.

The introduced here atomicity for the case of molecular system (corresponding to Mott-Hubbard localization effects in periodic systems) amounts to specifying a gradual transformation from molecular to atomic language in describing their 
electronic states, as a function of interatomic distance. This changeover is the basic feature and is associated with the essential change in regarding those particles as evolving within \emph{indistinguishable} (molecular) character and acquiring eventually the form of \emph{distinguishable} (atomic) states.

One must also underline that the concept of atomicity here is quantitative in nature. This is because the 
Mott-Hubbard localization concept in condensed-matter systems \cite{PRLSpal, Mott} appears usually as a first-order 
transition, requiring the energy equality of the two macro configuration (delocalized, localized) at this phase transition. Here the evolution may be regarded as a supercritical behavior at best \cite{PRLSpal, Sci, SpalA}.  However, the antiferromagnetic kinetic 
exchange survives even when the states are becoming orbitally distinguishable \cite{unpub}. 

Certainly, a further insight is required to quantify the present discussion for more complex systems. The present concepts are proposed to clarify 
the obviously \emph{unphysical behavior} of the increasing covalency with the increasing interatomic distance. As far as we are aware of, this inconsistency, although intuitively understandable, has not been discussed explicitly in the quantum-chemical literature. Also, the emerging atomicity here squares well with the Mott's original argument \cite{Mott} that the metallic (covalent) state of electrons in a periodic system is ruled out at (semi)macroscopic interatomic distances.

\section{Acknowledgment}

This work was supported by Grants OPUS No.~UMO-2018/29/B/ST3/02646 and No.~UMO-2021/41/B/ST3/04070 from Narodowe Centrum Nauki. We would like to thank Prof. Ewa ~Broc\l{}awik and Dr. Mariusz ~Rado\'n for discussions and criticism. We are also grateful to Andrzej Biborski and Andrzej ~P. Kadzielawa for making available to us their QMT library.

\appendix

\clearpage
\newpage
\section{Eigenvalues and eigenstates for \ch{H2} molecule}
\label{appendix:h2_molecule}

Starting from the orthogonalized restricted basis, we define the field operators as 

\begin{align}
    \hat{\psi}_{\sigma}(\textbf{r}) = w_{1}(\textbf{r})\chi_{\sigma}(1)\hat{a}_{1\sigma} +
    w_{2}(\textbf{r})\chi_{\sigma}(2)\hat{a}_{2\sigma}, \\ 
    \hat{\psi}^{\dagger}_{\sigma}(\textbf{r}) = w_{1}^{*}(\textbf{r})\chi_{\sigma}(1)\hat{a}^{\dagger}_{1\sigma} +
    w_{2}^{*}(\textbf{r})\chi_{\sigma}(2)\hat{a}^{\dagger}_{2\sigma}, 
\end{align}

\noindent
or, in compact notation, as

\begin{align}
\hat{\psi}(\textbf{r}) \equiv \begin{pmatrix}
\hat{\psi}_{\uparrow}(\textbf{r}) \\ \hat{\psi}_{\downarrow}(\textbf{r})
\end{pmatrix}.
  \label{eq:field_operators}
\end{align}

\noindent
In the above, $\hat{a}_{i\sigma}$ and $\hat{a}^{\dagger}_{i\sigma}$ are electron annihilation and creation operators in the state $w_{i\sigma}(\textbf{r}) \equiv w_1(\textbf{r})\chi_{\sigma}(1)$. Also, as we restrict here to $s$-orbital systems, the molecular (Wannier) functions can be taken as real if the condition $J'=J^H$ holds. Using the representation \eqref{eq:field_operators}, we obtain Hamiltonian~\eqref{Hamiltonian_eq} 
with the microscopic parameters expressed through the Slater orbitals and coefficients $\beta$ and 
$\gamma$ (cf. Eq.~\eqref{eq:beta_and_gamma}), or explicitly through inverse orbital size $\alpha$ and interatomic distance $R$ (see e.g. \cite{JS2000, JPCM2007}). The relevant physical quantities may be thus obtained as a function of $R$, with the orbital parameter $\alpha$ optimized in each case. 

To obtain the ground state energy $E_G$ for fixed $R$, the Hamiltonian~\eqref{Hamiltonian_eq} is diagonalized. This is carried out by making use of the global symmetry respecting two-particle states, leading to block-diagonal many-body Hilbert space, with specified values of the total spin, $S$, and its $z$-component, $S^z$, as well with transposition antisymmetry preserved. In effect, one can start the basis of $\binom{4}{2} = 6$ following states

\begin{align}
\begin{cases}
    &\ket{1} = \hat{a}^{\dagger}_{1\uparrow}\hat{a}^{\dagger}_{2\uparrow} \ket{0},\\
    &\ket{2} = \hat{a}^{\dagger}_{1\downarrow}\hat{a}^{\dagger}_{2\downarrow}\ket{0},\\
    &\ket{3} = \frac{1}{\sqrt{2}}(
    \hat{a}^{\dagger}_{1\uparrow}\hat{a}^{\dagger}_{2\downarrow}+\hat{a}^{\dagger}_{1\downarrow}\hat{a}^{\dagger}_{2\uparrow})\ket{0},\\
    &\ket{4} = \frac{1}{\sqrt{2}}(
    \hat{a}^{\dagger}_{1\uparrow}\hat{a}^{\dagger}_{2\downarrow}-\hat{a}^{\dagger}_{1\downarrow}\hat{a}^{\dagger}_{2\uparrow})\ket{0},\\
    &\ket{5} = \frac{1}{\sqrt{2}}(
    \hat{a}^{\dagger}_{1\uparrow}\hat{a}^{\dagger}_{1\downarrow}+\hat{a}^{\dagger}_{2\downarrow}\hat{a}^{\dagger}_{2\uparrow})\ket{0},\\
    &\ket{6} = \frac{1}{\sqrt{2}}(
    \hat{a}^{\dagger}_{1\uparrow}\hat{a}^{\dagger}_{1\downarrow}-\hat{a}^{\dagger}_{2\downarrow}\hat{a}^{\dagger}_{2\uparrow})\ket{0}.
    \end{cases}
\end{align}

\noindent
The first three are the spin-triplet states with $S^z=+1, -1, 0$, whereas the next three are inter- and intra-site singlets, respectively. The triplet state does not hybridize with other states and provides three $1 \times 1$ irreducible blocks with eigenvalues $\lambda_1=\lambda_2=\lambda_3 = \epsilon_1 + \epsilon_2 + K - J^H$. The remaining three singlet states compose the $3 \times 3$ block, so the Hamiltonian in that Fock subspace takes the form

\begin{align}
\hat{\mathcal{H}} =
    \begin{pmatrix}
    \epsilon + K + J^H & 2(t+V)    & 0 \\
    2(t+V)  & 2\epsilon + J + U   & \frac{1}{2}(U_1-U_2) \\ 
    0   & \frac{1}{2}(U_1-U_2) & 2\epsilon + U - J^H
    \end{pmatrix},
    \label{Hammatrix}
\end{align}

\noindent
where $\epsilon \equiv (\epsilon_1+\epsilon_2)/2$ and $U \equiv (U_1+U_2)/2$. This formulation allows to apply this formalism to both \ch{H2} (where $U_1 = U_2 = U$ and $\epsilon_1 +\epsilon_2 = \epsilon$), and to \ch{HeH+} and \ch{LiH}, where those simplifications are not met due to inequivalent atoms involved.

In the case of \ch{H2}, the eigenvalues of Eq.~\eqref{Hammatrix} take the form 

\begin{align}
    \lambda_{4,5} =& 2\epsilon + \frac{1}{2}(K+U) + J \pm \frac{1}{2} D, \\ 
    \lambda_6 =& 2\epsilon + U - J,
\end{align}

\noindent
with $D \equiv [(U-K)^2+16(t+V)^2]^{\frac{1}{2}}$. The corresponding eigenstates are 
\begin{align}
    \ket{\lambda_{4,5}} \equiv \ket{\lambda_{\pm}} = \frac{1}{[D(D\pm U\mp K)]^{\frac{1}{2}}} [(4(t+V))\ket{4} \pm (D \pm U \mp K)\ket{5}],
\end{align}

\noindent
where, for simplicity, we have defined the atomic-limit energy as the reference point, $\epsilon = 0$. We note that the eigenstates $\ket{\lambda_{4,5}}$ are superposed of the symmetric ionic state $\ket{5}$ and covalent part $\ket{4}$. The state $\ket{\lambda_5}$ is the ground state as the $\lambda_5$ eigenvalue is the lowest one. In the limit $U \gg |t+V|$, the $\lambda_5$ eigenvalue reads

\begin{align}
  \lambda_5 \simeq 2\epsilon + J^H + K - \frac{4(t+V)^2}{U-K}.
  \label{eq:lamgda_5_eval}
\end{align}

\noindent
The last term on the right-hand side of Eq.~\eqref{eq:lamgda_5_eval} is the so-called kinetic-exchange contribution. It competes with  ferromagnetic Heisenberg exchange $\sim J^H$. In similar manner, the two-particle states for \ch{HeH+} and \ch{LiH} are obtained, except that in those two cases, the diagonalization of the Hamiltonian matrix~\eqref{Hammatrix} cannot be carried out analytically, since the $\epsilon_1 \neq \epsilon_2$ and $U_1 \neq U_2$. The singlet state $\ket{\lambda_5}$ is elaborated further throughout the main text.

\clearpage
\newpage
\section{Tables of relevant quantities and parameters for considered systems}
\label{appendix:tables}

In Tables~\ref{table:params_h2}-\ref{table:params_lih} we provide relevant quantities and microscopic parameters versus $R$, obtained within 
EDABI scheme for the three systems discussed in main text, i.e., \ch{H2}, \ch{HeH+}, and \ch{LiH}.

\begin{table}[htbp]
  \caption{Ground state energy and microscopic parameters for H$_2$ molecule (in eV).}
  \label{table:params_h2}
\centering
\begin{tabular}{|c|l|l|l|l|l|l|l|}
\hline
$R$ ($a_0$)  & $E_G/N$ & $\epsilon$ & $t$ & $U$ & $K$ & $J$ & $V$ \\ \hline
0.5 & -6,57  & -14.29 & -31.75 & 30.06 & 19.43 & 0.43 & -0.28 \\ \hline
1   & -14,86 & -22.51 & -15.95 & 25.29 & 15.43 & 0.36 & -0.19 \\ \hline
1.5 & -15,58 & -23.84 & -9.21  & 22.08 & 12.67 & 0.29 & -0.16 \\ \hline
2   & -15,16 & -23.41 & -5.79  & 19.96 & 10.75 & 0.23 & -0.16 \\ \hline
2.5 & -14,61 & -22.56 & -3.84  & 18.61 & 9.34  & 0.18 & -0.16 \\ \hline
3   & -14,18 & -21.67 & -2.62  & 17.81 & 8.24  & 0.13 & -0.16 \\ \hline
3.5 & -13,89 & -20.86 & -1.82  & 17.38 & 7.35  & 0.09 & -0.16 \\ \hline
4   & -13,73 & -20.15 & -1.26  & 17.18 & 6.60  & 0.06 & -0.15 \\ \hline
4.5 & -13,65 & -19.52 & -0.86  & 17.09 & 5.95  & 0.04 & -0.14 \\ \hline
5   & -13,62 & -18.98 & -0.59  & 17.05 & 5.40  & 0.02 & -0.12 \\ \hline
\end{tabular}
\end{table}

Note that the value of $|t|$ is comparable to U in the limit $R<R_{bond}$ and diminishes spectacularly when $R>R_{bond}$ (i.e. in the strong-correlation regime).

\begin{table}[htbp]
  \caption{Ground state energy and microscopic parameters for \ch{HeH+} molecular ion (in eV).}
  \label{table:params_heh+}
\centering
\begin{tabular}{|c|l|l|l|l|l|l|l|l|}
\hline
$R$ ($a_0$)  & $E_G/N$ & $\epsilon_\mathrm{H}$ & $\epsilon_{\mathrm{He}}$ & t & $U_\mathrm{H}$ & $U_{\mathrm{He}}$ & $K$ & $V$ \\ \hline
0.5 & -27.84 & -22.42 & -14.32 & -24.07 & 22.35 & 36.70 & 11.17 & -0.99 \\ \hline
1   & -38.54 & -32.81 & -32.43 & -18.96 & 20.57 & 21.85 & 8.48  & -0.75 \\ \hline
1.5 & -39.84 & -34.59 & -33.13 & -15.49 & 19.54 & 14.62 & 6.72  & -0.60 \\ \hline
2   & -39.63 & -34.10 & -29.57 & -13.15 & 18.95 & 11.10 & 5.66 & -0.50 \\ \hline
2.5 & -39.38 & -32.20 & -25.93 & -11.55 & 18.61 & 9.39  & 4.95  & -0.44 \\ \hline
3   & -39.21 & -29.67 & -22.86 & -10.38 & 18.41 & 8.56  & 4.54  & -0.39 \\ \hline
3.5 & -37.68 & -27.20 & -20.45 & -9.45  & 18.30 & 8.15  & 4.26  & -0.36 \\ \hline
4   & -39.09 & -25.08 & -18.62 & -8.64  & 18.23 & 7.96  & 4.05  & -0.33 \\ \hline
4.5 & -38.97 & -23.37 & -17.23 & -7.90  & 18.19 & 7.86  & 3.86  & -0.31 \\ \hline
5   & -38.92 & -22.01 & -16.16 & -7.20  & 18.17 & 7.82  & 3.69  & -0.31 \\ \hline
\end{tabular}
\end{table}

\begin{table}[htbp]
\centering
\caption{Ground state energy and microscopic parameters for \ch{LiH} molecule (in eV).}
  \label{table:params_lih}
\begin{tabular}{|c|c|c|c|c|c|c|c|c|}
\hline
$R$ ($a_0$)  & $E_G/N$ & $\epsilon_\mathrm{H}$ & $\epsilon_{2s\mathrm{Li}}$ & $t$ & $U_\mathrm{H}$ & $U_{2s\mathrm{Li}}$ &  $K$ & $V$ \\ \hline
1   & -98.21  & -43.25   & -39.32   &  -55.89 &   38.29  &  33.08 & 21.19  & -1.70 \\ \hline
1.5 & -105.89 & -44.47   & -40.24   &  -48.02 &   35.13  &  26.34 & 19.79  & -1.52 \\ \hline
2   & -108.10 & -45.402  & -40.97   &  -37.97 &   32.02  &  22.01 & 17.10  & -1.02 \\ \hline
2.5 & -108.88 & -45.73   & -41.97   &  -30.93 &   28.89  &  17.98 & 16.77  & -0.89 \\ \hline
3   & -109.29 & -46.12   & -42.53   &  -24.43 &   24.80  &  14.04 & 15.12  & -0.78 \\ \hline
3.5 & -109.35 & -45.93   & -41.55   &  -19.89 &   22.08  &  11.35 & 13.29  & -0.69 \\ \hline
4   & -109.29 & -45.81   & -41.53   &  -14.05 &   20.30  &   9.44 & 11.98  & -0.51 \\ \hline
4.5 & -109.04 & -45.50   & -41.44   &  -11.23 &   19.19  &   8.41 & 10.13  & -0.38 \\ \hline
5   & -108.92 & -45.75   & -41.39   &  -8.09  &   18.09  &   7.75 &  9.48  & -0.32 \\ \hline
5.5 & -108.71 & -44.96   & -40.91   &  -5.48  &   18.25  &   7.48 &  6.32  & -0.28 \\ \hline
6   & -108.32 & -44.49   & -40.76   &  -4.01  &   17.98  &   7.25 &  4.01  & -0.26 \\ \hline
\end{tabular}
\end{table}

The RHF and CI computations were carried out using the GAMESS code and the 6-31G basis set to represent the Slater functions. Numerical accuracy for the EDABI calculations is $10^{-4}a_0$ for R and $10^{-5}eV$ for energy, respectively. 

\clearpage
\newpage
\printbibliography
\end{document}